\pdfoutput=1
\documentclass[letterpaper,10pt]{article}
\usepackage[utf8]{inputenc}

\usepackage[english]{babel}
\usepackage{csquotes}

\usepackage{amsfonts}
\usepackage{amsmath}
\usepackage{amssymb}
\usepackage{amsthm}
\usepackage[backend=biber,style=numeric]{biblatex}
\usepackage{geometry}
\usepackage{subcaption}
\usepackage{tikz}
\usepackage{xcolor}

\usepackage{hyperref}

\usetikzlibrary{backgrounds}
\usetikzlibrary{calc}
\usetikzlibrary{positioning}
\usetikzlibrary{shapes}

\theoremstyle{definition}
\newtheorem{definition}{Definition}[section]

\begin{document}

\title{Building Scalable Decentralized Payment Systems}
\author{
John Adler \\
\href{mailto:john.adler@protonmail.ch}{\nolinkurl{john.adler@protonmail.ch}} \\
\and
Mikerah Quintyne-Collins \\
\href{mailto:mikerah@hashcloak.com}{\nolinkurl{mikerah@hashcloak.com}} \\
}
\date{\today}

\maketitle

\begin{center}
\parbox{0.8\linewidth}{
\noindent \textbf{Abstract.}
Increasing the transactional throughput of decentralized blockchains in a secure manner has been the holy grail of blockchain research for most of the past decade.
This paper introduces a scheme for scaling blockchains while retaining virtually identical security and decentralization, colloquially known as optimistic rollup.
We propose a layer-2 scaling technique using a permissionless side chain with merged consensus.
The side chain only supports functionality to transact UTXOs and transfer funds to and from a parent chain in a trust-minimized manner.
Optimized implementation and engineering of client code, along with improvements to block propagation efficiency versus currently deployed systems, allow use of this side chain to scale well beyond the capacities exhibited by contemporary blockchains without undue resource demands on full nodes.
}
\end{center}

\section{Introduction}
\label{sec:intro}

The introduction of decentralized blockchains, initially conceived as a means for cash payments without a trusted intermediary in the form of Bitcoin~\cite{bitcoin}, has sparked a flurry of interest in developing decentralized applications for a wide variety of areas.
Smart contracts~\cite{szabo1997}, programs whose consistent global execution is enforced by a consensus protocol among a decentralized network of nodes rather than by a single server, have in recent years gained traction as a means of dis-intermediating a wide assortment of non-financial tasks.

Unfortunately, the base layers of fully decentralized blockchain systems, as deployed presently, are extremely limited in their transactional throughput.
The Bitcoin blockchain currently processes an average of $\sim$3 transactions per second~\cite{scaling} and is operating at maximum capacity, while the Ethereum blockchain is currently capped at $\sim$15 transactions per second and is often operating at its maximum capacity as well.
In contrast, global payment processors handle on the order of tens of thousands of transactions per second~\cite{scaling}.

This work presents a study on numerous scaling methodologies devised over the past decade and analyzes their features and challenges.
A novel scaling direction that is mostly composed of well-known and studied components is then introduced~\footnote{An earlier version of this work introduced the high-level ideas that were later refined into a minimal viable spec~\cite{minimal_viable_merged_consensus} of what is now known as ``optimistic rollup~\cite{optimistic_rollup_pg,fuel_labs}.'' The present version collects the minimal spec and subsequent improvements into a single cohesive document. A layperson's edition of this work is also available~\cite{the_whys_of_oru}.}.
A side chain construction is used to avoid any mainnet protocol changes, which would require coordinating a fork with client developers, users, application developers, and exchanges, while allowing innovations and improvements to be deployed~\cite{sidechains,forks}.
Merged consensus is used to progress the side chain, which borrows security from the parent chain using a commit chain scheme~\cite{nocust}.
Finally, only a bare minimum of functionality is enabled on the side chain, allowing only financial transactions and trust-minimized movement of funds between the side chain and its parent chain.

Section~\ref{sec:prelim} presents fundamental technical preliminaries.
Section~\ref{sec:method} describes our proposal for scaling using side chains with merged consensus.
Finally, section~\ref{sec:related} gives an overview of and contrasts previous scaling proposals.

\section{Preliminaries}
\label{sec:prelim}

This section gives a high-level overview of the fundamental techniques that will be built upon for our proposed scaling solution.
It also provides more precise definitions for various terms whose use is widespread but are often poorly, incorrectly, or incompletely defined.

\subsection{Decentralized Blockchains}
\label{sec:prelim:blockchain}

\begin{figure}[!h]
\centering
\begin{tikzpicture}[every node/.style = {shape=rectangle, draw=black, align=center, minimum size=1cm}]
\node (b1) [draw opacity=0] {\dots};
\node (b2) [right=of b1] {$B_{i-2}$};
\node (b3) [right=of b2] {$B_{i-1}$};
\node (b4) [right=of b3] {$B_{i}$};
\path[->]
    (b2) edge (b1)
    (b3) edge (b2)
    (b4) edge (b3)
    ;
\end{tikzpicture}
\caption{Blockchain structure.}
\label{fig:blockchain}
\end{figure}
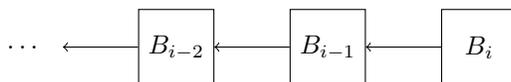

At its core, a \textit{blockchain} is nothing more than a database consisting of a cryptographic-hash-linked chain of blocks that defines a total ordering of transactions and is deterministically verifiable.
A blockchain is deterministically verifiable if its correctness can be determined using only data contained within itself (\textit{i.e.}, it is self-consistent), and is accomplished through the use of cryptographic hashing to link blocks together and digital signatures for each transaction.
Execution engines can be built on top of this ordered (\textit{i.e.}, serialized) data, such as the Ethereum Virtual Machine (EVM) in Ethereum~\cite{ethereum}.
Pictured in Figure~\ref{fig:blockchain} is a high-level representation of the blockchain data structure, with each block (square) containing a hash of the previous block in the chain (arrow).

\textit{Decentralized} blockchains are of particular interest for use in systems with open participation that must be publicly auditable, such as for a dis-intermediated payments system.

\begin{definition}
A blockchain system is \textit{distributed} if it is replicated across \textbf{more than one} physical computer (a system participant, \textit{i.e.}, a node).
\end{definition}

\begin{definition}
\label{def:permissionless}
A blockchain system is \textit{permissionless} if it can be read from and written to without requiring permission from \textbf{existing} system participants.
In practical terms, this property requires users to be able to participate as block producers of the system without first being part of the system.
\end{definition}

Colloquially, a blockchain system is \textit{trustless} if it does not require trusting \textbf{any} external resources to interact with it (including, but not limited to, a third-party computer, an escrow service, a trusted notary, and binary executables with no source code or non-deterministic builds).
This description of ``trustless'' is circular however, as ``trust'' is not defined; to that end, let us examine presently a more rigorous definition of the word.

State elements in a blockchain system come in two variants: owned (associated with a public address), or unowned (not associated with any address \textit{i.e.}, unused).
Delegation of ownership is possible, so the ``owner'' of a state element can be considered both the owner of the private key associated with the address of the owned state element and any potential delegated owners that can be granted varying permissions over the state element.

\begin{definition}
An owned state element is \textit{live} if it can be modified by its owner (or by its delegated owners within their permissions) in finite time.
An unowned state element is live if it can become owned in finite time.
\end{definition}

\begin{definition}
An owned state element is \textit{safe} if it can never be modified by any of its non-owners (or by its delegated owners outside their permissions).
Unowned state elements are trivially safe.
\end{definition}

We can use these to form a concrete, and more importantly non-circular, definition of ``trustlessness'':

\begin{definition}
\label{def:trustless_concrete}
A blockchain system is \textit{trustless} if and only if its state is (\textit{i.e.}, all its state elements are) both live and safe.
\end{definition}

\noindent
We now compose the previous definitions to arrive at a useable and useful definition of the term ``decentralized.''

\begin{definition}
\label{def:decentralized}
A blockchain system is \textit{decentralized} if and only if it is 1) distributed and 2) trustless and 3) permissionless.
\end{definition}

Definition~\ref{def:trustless_concrete} is useful when evaluating layer-2 scaling techniques.
Unlike layer-1 blockchain systems, which are physical systems, layer-2 constructions that anchor onto parent chains for security and other guarantees can be thought of as logical abstractions, for which the notion of trust in physical machines or persons isn't useful.

It is trivial to see that a blockchain system that is trustless by Definition~\ref{def:trustless_concrete} under no assumptions will violate the FLP impossibility~\cite{flp}.
Indeed, even layer-1 blockchain systems are only trustless under a majority block producer assumption: in the case of PoW blockchains, a majority of miners can censor transactions indefinitely, violating state liveness.
Therefore our goal is to minimize the assumptions needed to make such systems trustless.

\subsection{Consensus Protocols}
\label{sec:prelim:consensus}

Permissionless blockchains (see \S~\ref{sec:prelim:blockchain} for definitions), require a consensus protocol for writing blocks to the database~\cite{consensus}.
Given their permissionless nature, they require some form of Sybil resistance mechanism (\textit{i.e.}, impersonating multiple users should not grant more power in the protocol).
Nakamoto Consensus, introduced for use in Bitcoin~\cite{bitcoin}, is the first consensus protocol that performs in a permissionless setting, and leverages Proof-of-Work (PoW) as its Sybil resistance mechanism.
A cryptographic hash function can be used, modeled as a random oracle, to determine a block producer~\cite{backbone}, with each participant having a chance of being a block producer proportional to the computational power they devote to the protocol.
The \textit{longest chain} (or, more precisely, heaviest chain) of valid blocks, each with sufficient proofs of work, is considered the canonical chain---this is the fork choice rule generally used by the family of consensus protocols based on Nakamoto Consensus.

The primary function of these consensus protocols is to provide \textit{security} to the chain.
\begin{definition}
The \textit{security} of a blockchain system is the cost of changing its history (\textit{i.e.}, rewriting blocks through a chain re-organization).
\end{definition}
\noindent
This concept of blockchain security is distinct from \textit{e.g.}, cryptographic security or smart contract security.

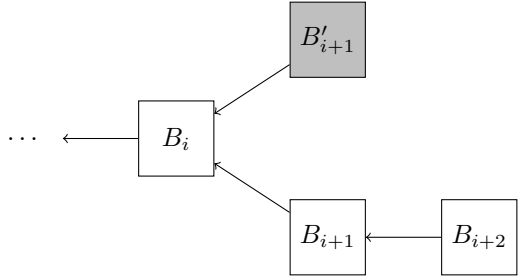
\begin{figure}[!h]
\centering
\begin{tikzpicture}[every node/.style = {shape=rectangle, draw=black, align=center, minimum size=1cm}]
\node (i) [draw opacity=0] {\dots};
\node (a1) [right=of i] {$B_{i}$};
\node (b1) [above right=0.3cm and 1cm of a1, fill=lightgray] {$B'_{i+1}$};
\node (c1) [below right=0.3cm and 1cm of a1] {$B_{i+1}$};
\node (c2) [right=of c1] {$B_{i+2}$};
\path[->]
    (a1) edge (i)
    (b1) edge (a1)
    (c1) edge (a1)
    (c2) edge (c1)
    ;
\end{tikzpicture}
\caption{Blockchain reorganization begins. Previously-shorter chain is marked in gray.}
\label{fig:reorg_begin}
\end{figure}

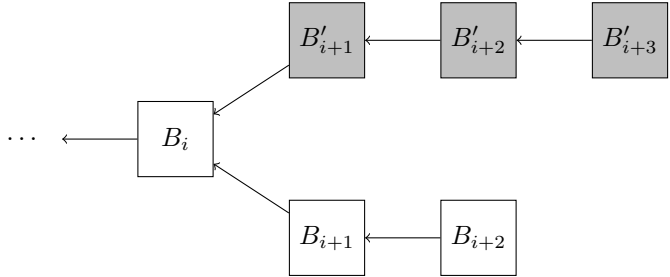
\begin{figure}[!h]
\centering
\begin{tikzpicture}[every node/.style = {shape=rectangle, draw=black, align=center, minimum size=1cm}]
\node (i) [draw opacity=0] {\dots};
\node (a1) [right=of i] {$B_{i}$};
\node (b1) [above right=0.3cm and 1cm of a1, fill=lightgray] {$B'_{i+1}$};
\node (b2) [right=of b1, fill=lightgray] {$B'_{i+2}$};
\node (b3) [right=of b2, fill=lightgray] {$B'_{i+3}$};
\node (c1) [below right=0.3cm and 1cm of a1] {$B_{i+1}$};
\node (c2) [right=of c1] {$B_{i+2}$};
\path[->]
    (a1) edge (i)
    (b1) edge (a1)
    (b2) edge (b1)
    (b3) edge (b2)
    (c1) edge (a1)
    (c2) edge (c1)
    ;
\end{tikzpicture}
\caption{Blockchain reorganization complete. Previously-shorter chain is now longer.}
\label{fig:reorg_complete}
\end{figure}

Concerns over the shortcomings of Nakamoto Consensus-style protocols (lack of strong finality guarantees, shown in Figures~\ref{fig:reorg_begin} and~\ref{fig:reorg_complete}, where a previously-shorter chain overtakes the previously-longest chain and becomes the new canonical chain) along with the continued use of PoW (enormous energy waste) have led to the search of new consensus protocols that employ stake-based Sybil-resistance, known as Proof-of-Stake (PoS)~\cite{ethereum,ouroboros,avalanche2}.

The replication of the many desirable properties of Nakamoto Consensus, and the few undesirable properties, with stake-based consensus protocols has been unsuccessful, however.
No blockchain system employing a stake-based Sybil-resistant permissionless consensus protocol with satisfactory properties has been deployed in practice as of this writing~\cite{longest_chain_pos}, and to the best of our knowledge, despite many claims to the contrary no such system has been devised to date.
In addition, purely stake-based consensus protocols are not decentralized by Definition~\ref{def:decentralized}---specifically Definition~\ref{def:permissionless}---as 1) there is no known way to fairly distribute stake initially and 2) participation in the system requires coins or tokens to be purchased from system participants.
Both of these problems are solved by Nakamoto Consensus' use of PoW.

\subsection{The Scaling Problem}
\label{sec:prelim:scalingproblem}

Limitations in transactional throughput for public blockchains, colloquially known as ``The Scaling Problem,'' present a significant roadblock to real-world adoption of such systems.
The root cause of the scaling bottleneck is that every block in a decentralized blockchain network must be fully validated by every node (client) on the network.
Transaction throughput can be increased trivially by sacrificing security or decentralization, so the true challenge lies in designing a system that is scalable, secure, and decentralized.

Almost universally, scaling proposals aim to not have every node validate every block, but rather have a subset of nodes validate a subset of (relevant, in some way) transactions.
Layer-1 scaling proposals aim to increase the transaction throughput of the base chain, and generally employ \textit{sharding}~\cite{scaling}, splitting up transactions and state into individual shards instead of collecting them all into a single logical chain.
In this model, transaction throughput is increased proportionally to the number of shards (minus overhead for managing shards).
It should be noted that PoW-based sharding is undesirable, as security would be split between shards.

Layer-2 scaling proposals~\cite{sidechains,state_channels,counterfactual,plasma,lighting_network,arbitrum} aim to move groups of transactions off-chain (or, more precisely, away from the parent chain, \textit{e.g.}, Bitcoin or Ethereum, and onto a second layer network).
Transactions can be grouped by type, or application.
For example, micropayments can be done through a payment channel network~\cite{lighting_network}, or transactions specific to a single application can be processed through their own chain~\cite{plasma}.

We shall see later that the scaling solution proposed in this work is also centered around the idea of not having every node on the network validate every transaction, namely by separating execution (validation) from data ordering availability.

\subsection{Improvements to Block Propagation}
\label{sec:prelim:propagation}

As mentioned in \S~\ref{sec:prelim:scalingproblem}, the scaling bottleneck is due to every node fully validating every block.
A prerequisite to validating a new block when one is produced by the network is downloading it in its entirety.

To this end, several techniques have been proposed~\cite{bip152,graphene,minisketch}, based largely on set reconciliation using bloom filters~\cite{bloom}, invertible bloom lookup tables~\cite{iblt}, and sketches~\cite{sketches}, to allow blocks to be constructed locally based on an extremely compressed representation of a block's included transactions.
Using these, block-producing nodes can spread out their network bandwidth requirements over the entire length of the blocktime, downloading transactions as they are propagated through the network only once.

\subsection{Fraud Proofs and Data Availability}
\label{sec:prelim:fraudproofs}

A model for generalized fraud proofs introduced in~\cite{fraud_proofs} allows for trust-minimized light clients.
Non-fully-validating nodes (known as light nodes, or light clients), only check block headers for validity---in a PoW blockchain, that valid and sufficient proof of work was done.
The contents of blocks must be assumed to be too expensive for a light client to ever download and validate for even a single block.

The proposed fraud proof scheme modifies the transaction Merkle tree to add intermediate state root commitments into it.
A fraud proof can then consist of a parametrizable number of Merkle branches against the initial (possibly intermediate) pre-state from which to begin applying transactions, and comparing the resultant state root with the committed post-state root.

In addition to fraud proofs,~\cite{fraud_proofs} proposes to use erasure codes~\cite{erasure_codes} for data availability proofs, which are needed as a fraud proof can't be generated for an unavailable block.

These data availability proofs involve erasure coding each block, with clients randomly sampling a fixed number of samples.
A fraud proof and associated synchrony assumption is needed in case the erasure coding was performed incorrectly.
Using a 2D erasure coding scheme, fraud proofs are at most $O(\sqrt{n})$ cost (where $n$ is the blocksize).
A later work~\cite{coded_merkle_tree} proposed an order-optimal variant of this scheme, with $O(\log{n})$ fraud proof cost.

The existence of compact fraud proofs and data availability proofs allow light clients to operate with reduced trust assumptions.
Whereas without these proofs light clients required trust in a majority of block producers being honest, with these proofs light clients only require trust that a single honest node exists in the network that is capable of relaying proofs to them.
In practice one will find that this trust assumption is not objectively stronger than the trust the vast majority of users place on, for example, the hardware manufacturer of their CPU, or the implementation and design of cryptographic hash functions without backdoors.

\section{Scaling Decentralized Blockchains}
\label{sec:method}

This section discusses in-depth our proposed scaling solution of a side chain with merged consensus for financial transactions.
The proposed construction is capable of handling a large number of transactions per second with virtually identical security and decentralization as its parent chain.
Comparisons to other scaling proposals are also discussed.

Without loss of generality, we will assume the parent chain of this system is Ethereum~\cite{ethereum} and use associated vocabulary.
Any chain with sufficient expressibility for smart contracts and statefulness will suffice.

\subsection{A Side Chain for Financial Transactions}
\label{sec:method:sidechain}

Despite suggestions to parallelize validation of blocks in Ethereum, the bottleneck of client software is in practice disk I/O bandwidth~\cite{eip648}.
A combination of poor design of the EVM's opcodes, complex expressivity, and use of an inefficient state trie data structure make it challenging to develop both efficient software and potentially hardware to validate transactions.

This work proposes a side chain construction with just enough expressivity for performing financial transactions and trust-minimized movement of funds between the side chain and its parent chain (with optional stateless predicate scripting functionality to support a subset of state channel~\cite{state_channels} and other constructions).
Rather than the accounts data model of the EVM, a UTXO data model is used, as the latter is simpler to reason about and optimize parallel implementations for in practice.

The side chain's consensus protocol is \textit{merged consensus}, a permissionless consensus protocol that runs entirely on-chain.
This is discussed in more detail in Section~\ref{sec:method:merged_consensus}.
Security is borrowed from the main chain by timestamping side chain blocks, which prevents history-rewriting attacks that do not also affect the main chain.

Thanks to the existence of general-purpose compact fraud proofs and data availability proofs (\S~\ref{sec:prelim:fraudproofs}), the number of transactions included per block can be increased to an arbitrarily large size bound only by physical limitations of block-producing (\textit{i.e.}, mining) nodes to transmit large quantities of data with low computational complexity, while the overall system still remains decentralized.
As with all other scalability proposals, increased transaction throughput is achieved by not having every node in the network validate every transaction: with this scheme, main chain block producers are only required to order the side chain's data, while the side chain is validated by its participants.

It should be noted that while the design presented in this section is most suited for building a scalable decentralized payment system, it can be trivially extended to support a general-purpose stateful smart contract execution platform (albeit with only part of the performance gains)~\cite{minimal_viable_merged_consensus}.

\subsection{Side Chain Design}
\label{sec:method:design}

This section gives a bird's-eye view of the proposed side chain design, which is then analyzed in subsequent sections.

A contract is deployed onto the parent chain that will keep track of side chain block headers, deposits and withdrawals, and process fraud proofs.
Using a leader selection protocol that runs entirely on-chain (discussed in \S~\ref{sec:method:merged_consensus}), the leader can submit a side chain block to the contract, along with a bond of parametrizable size.
This block \textit{must} extend the tip of the side chain known to the contract, otherwise it is immediately rejected.
The side chain block is posted in its entirety in the transaction's data field \textit{e.g.}, \texttt{calldata} for Ethereum~\cite{roll_up}.
The contract will authenticate (\textit{i.e.}, Merklelize) the side chain's transactions, and either compare this against the posted transactions root, or compute the actual block header hash (an implementation detail).
Finally, the block header hash is saved by the contract for later use---essentially, the contract runs a light client of the side chain.

After a parametrizable finalization delay, an unchallenged side chain block can be \textit{finalized} (\textit{i.e.}, becomes irreversible).
If a non-finalized side chain block includes an invalid state transition, anyone may submit a fraud proof~\cite{fraud_proofs} on-chain which, if valid, rolls back the tip of the side chain to the previous block and rewards half the bonds of orphaned side chain blocks to the prover (the other half is effectively or explicitly burned).

Side chain block producers and incentivized to fully validate the chain, lest they extend an invalid block and have their bond burned.

Deposits are trivial in this system: users can simply send funds to the contract then spend them on the side chain immediately.
Withdrawals can be accomplished by first burning funds on the side chain, then posting a non-interactive withdrawal request on-chain with an inclusion proof of this burn against a finalized side chain block.

Transaction latency can be as fast as the parent chain's block times: with client-side validation of the fully-available data, users (and side chain block producers) can convince themselves that a block is valid without having to wait until it is finalized.
As otherwise valid blocks that build upon a valid history are guaranteed to eventually finalized by construction, there is no additional latency introduced.

\subsection{Merged Consensus}
\label{sec:method:merged_consensus}

\textit{Merged consensus} is a consensus protocol that is fully verifiable on-chain.
While the name bears similarity to merged mining (\S~\ref{sec:related:mergedmining}), merged consensus provides us with vastly different properties.

Recall that decentralize consensus protocols give a blockchain security (\S~\ref{sec:prelim:consensus}).
These protocols usually consist of a number of distinct components:
\begin{enumerate}
\item
A fork choice rule: how to choose between two otherwise valid chain.

\item
A block validity function: completely defines the state transition.

\item
A leader selection algorithm: how to determine who gets to progress the chain by extending the tip with a new block. (Some decentralized consensus protocols are leaderless~\cite{avalanche2}.)

\item
A Sybil resistance mechanism: such as Proof-of-Work or Proof-of-Stake.
\end{enumerate}

The proposed side chain scheme is \textit{fork-free} by construction---as new blocks are enforced to only be able to extend the single tip---so only a trivial fork choice rule is needed.
The block validity function is a simple UTXO data model with optional stateless predicate scripting for scalable payments, and potentially any arbitrary computational model if that is desired.
This leaves leader selection and Sybil resistance.

\begin{figure}[!h]
\captionsetup{justification=centering}
\centering

\begin{tikzpicture}[every node/.style = {shape=rectangle, draw=black, align=center, minimum size=1cm}]
\node (idots) [draw opacity=0] {\dots};
\node (sdots) [draw opacity=0, above=0.4cm of idots] {\dots};
\node (i) [right=of idots] {$B_{i}$};
\node (s) [right=of sdots] {$S_{j}$};

\node (a1) [right=of i]  {$B_{i+1}$};
\node (a2) [right=of a1] {$B_{i+2}$};
\node (a3) [right=of a2] {$B_{i+3}$};
\node (a4) [right=of a3] {$B_{i+4}$};
\node (a5) [right=of a4] {$B_{i+5}$};
\node (a6) [right=of a5] {$B_{i+6}$};

\node (s1) [draw opacity=0, right=of s] {};
\node (s2) [right=of s1] {$S_{j+1}$};
\node (s3) [right=of s2] {$S_{j+2}$};
\node (s4) [draw opacity=0, right=of s3] {};
\node (s5) [right=of s4] {$S_{j+3}$};
\node (s6) [right=of s5] {$S_{j+4}$};
\path[->]
    (i) edge (idots)
    (s) edge (sdots)
    (i) edge (s)
    (a1) edge (i)
    (a2) edge (a1)
    (a2) edge (s2)
    (a3) edge (a2)
    (a3) edge (s3)
    (a4) edge (a3)
    (a5) edge (a4)
    (a5) edge (s5)
    (a6) edge (a5)
    (a6) edge (s6)
    ;
\path[->]
    (s2) edge (s)
    (s3) edge (s2)
    (s5) edge (s3)
    (s6) edge (s5)
    ;
\end{tikzpicture}

\caption{Example normal operation of side chain with merged consensus.}
\label{fig:sc_merged_consensus}
\end{figure}
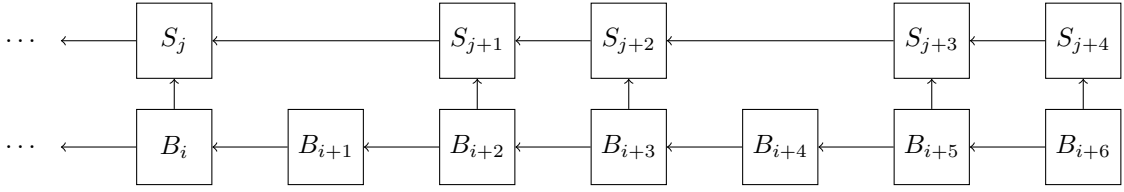

We propose the simplest possible leader selection algorithm: ``first come, first served''~\cite{minimal_viable_merged_consensus}.
The first transaction that gets included on the parent chain that successfully extends the tip is the post-facto leader.
Sybil resistance is provided by parent chain transaction fees and inherent rate-limiting.
Normal operation of the side chain and parent chain is shown in Figure~\ref{fig:sc_merged_consensus}; note that unlike in merged mining (Figure~\ref{fig:merged_mining}) side chain blocks cannot be produced without a linked parent chain block.

This extremely simple proposal works because \textit{the parent chain already provides security} in the form of a timestamping server for non-repudiation and non-equivocation~\cite{catena}.
This is known as the commit chain paradigm.

\begin{definition}
A \textit{commit chain}~\cite{nocust} is a side chain that borrows security from its parent chain through periodic commitment of block hashes (\textit{i.e.}, including a side chain block hash into the parent chain as a state transition).
\end{definition}

If a more orderly leader selection algorithm is desired (say, for easier block propagation), anything that runs entirely on-chain can be used, such as randomly shuffling staked validators using an on-chain random number generator~\cite{randao}.
Staking in this manner does not need a separate token, as the whole purpose of a native coin as originally envisioned by Nakamoto was to provide a disincentive against a majority-hashrate history rewrite~\cite{bitcoin}---a non-issue with merged consensus as the parent chain provides security.

\begin{figure}[!h]
\captionsetup{justification=centering}
\centering

\begin{subfigure}[b]{0.4\textwidth}
\begin{tikzpicture}[every node/.style = {shape=rectangle, draw=black, align=center, minimum size=1cm}]
\node (idots) [draw opacity=0] {\dots};
\node (sdots) [draw opacity=0, above=0.4cm of idots] {\dots};
\node (i) [right=0.5cm of idots, fill=red!50] {$B_{i}$};
\node (s) [right=0.5cm of sdots, fill=red!50] {$S_{j}$};

\node (a1) [above right=0.5cm and 0.5cm of i, fill=red!50] {$B^{A}_{i+1}$};
\node (a2) [right=0.5cm of a1, fill=red!50] {$B^{A}_{i+2}$};
\node (s1) [above=0.4cm of a1, fill=red!50] {$S^{A}_{j+1}$};
\node (s2) [right=0.5cm of s1, fill=red!50] {$S^{A}_{j+2}$};

\node (b1) [below right=0.5cm and 0.5cm of i] {$B^{B}_{i+1}$};
\node (b2) [right=0.5cm of b1] {$B^{B}_{i+2}$};
\node (t1) [above=0.4cm of b1] {$S^{B}_{j+1}$};
\node (t2) [right=0.5cm of t1] {$S^{B}_{j+2}$};
\path[->]
    (i) edge (idots)
    (s) edge (sdots)
    (i) edge (s)
    (a1) edge (i)
    (a1) edge (s1)
    (a2) edge (a1)
    (a2) edge (s2)
    (b1) edge (i)
    (b1) edge (t1)
    (b2) edge (b1)
    (b2) edge (t2)
    ;
\path[->]
    (s1) edge (s)
    (s2) edge (s1)
    (t1) edge (s)
    (t2) edge (t1)
    ;
\end{tikzpicture}
\caption{Alice's view of the chain, in red.}
\end{subfigure}
\begin{subfigure}[b]{0.4\textwidth}
\begin{tikzpicture}[every node/.style = {shape=rectangle, draw=black, align=center, minimum size=1cm}]
\node (idots) [draw opacity=0] {\dots};
\node (sdots) [draw opacity=0, above=0.4cm of idots] {\dots};
\node (i) [right=0.5cm of idots, fill=blue!50] {$B_{i}$};
\node (s) [right=0.5cm of sdots, fill=blue!50] {$S_{j}$};

\node (a1) [above right=0.5cm and 0.5cm of i] {$B^{A}_{i+1}$};
\node (a2) [right=0.5cm of a1] {$B^{A}_{i+2}$};
\node (s1) [above=0.4cm of a1] {$S^{A}_{j+1}$};
\node (s2) [right=0.5cm of s1] {$S^{A}_{j+2}$};

\node (b1) [below right=0.5cm and 0.5cm of i, fill=blue!50] {$B^{B}_{i+1}$};
\node (b2) [right=0.5cm of b1, fill=blue!50] {$B^{B}_{i+2}$};
\node (t1) [above=0.4cm of b1, fill=blue!50] {$S^{B}_{j+1}$};
\node (t2) [right=0.5cm of t1, fill=blue!50] {$S^{B}_{j+2}$};
\path[->]
    (i) edge (idots)
    (s) edge (sdots)
    (i) edge (s)
    (a1) edge (i)
    (a1) edge (s1)
    (a2) edge (a1)
    (a2) edge (s2)
    (b1) edge (i)
    (b1) edge (t1)
    (b2) edge (b1)
    (b2) edge (t2)
    ;
\path[->]
    (s1) edge (s)
    (s2) edge (s1)
    (t1) edge (s)
    (t2) edge (t1)
    ;
\end{tikzpicture}
\caption{Bob's view of the chain, in blue.}
\end{subfigure}

\caption{Two miners, Alice $A$ and Bob $B$, mining on different heads of equal height.}
\label{fig:sc_reorg1}
\end{figure}
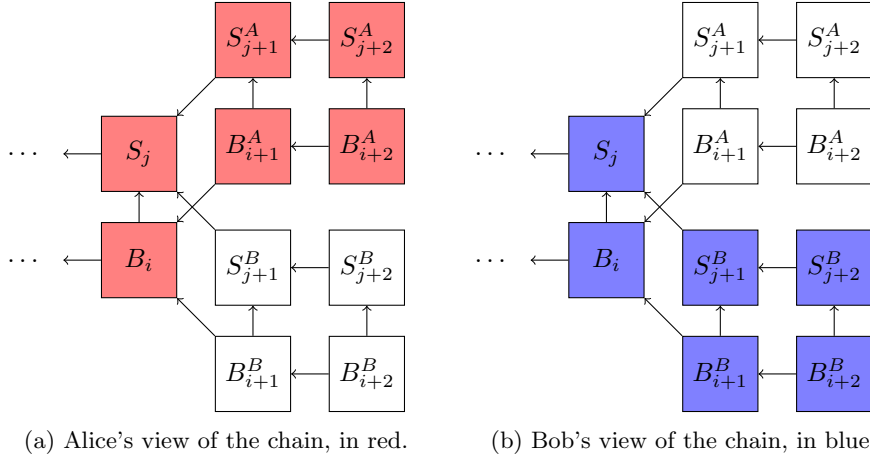

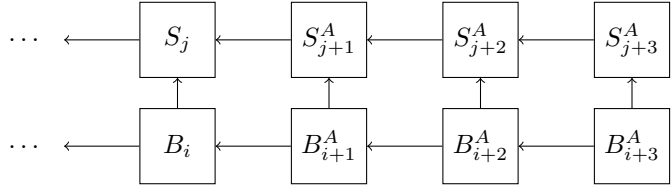
\begin{figure}[!h]
\captionsetup{justification=centering}
\centering
\begin{tikzpicture}[every node/.style = {shape=rectangle, draw=black, align=center, minimum size=1cm}]
\node (idots) [draw opacity=0] {\dots};
\node (sdots) [draw opacity=0, above=0.4cm of idots] {\dots};
\node (i) [right=of idots] {$B_{i}$};
\node (s) [right=of sdots] {$S_{j}$};

\node (a1) [right=of i]  {$B^{A}_{i+1}$};
\node (a2) [right=of a1] {$B^{A}_{i+2}$};
\node (a3) [right=of a2] {$B^{A}_{i+3}$};
\node (s1) [right=of s]  {$S^{A}_{j+1}$};
\node (s2) [right=of s1] {$S^{A}_{j+2}$};
\node (s3) [right=of s2] {$S^{A}_{j+3}$};
\path[->]
    (i) edge (idots)
    (s) edge (sdots)
    (i) edge (s)
    (a1) edge (i)
    (a1) edge (s1)
    (a2) edge (a1)
    (a2) edge (s2)
    (a3) edge (a2)
    (a3) edge (s3)
    ;
\path[->]
    (s1) edge (s)
    (s2) edge (s1)
    (s3) edge (s2)
    ;
\end{tikzpicture}
\caption{Alice finds the next parent chain block, so Bob reorganizes his local chain to follow the longest chain he knows of. Both Alice and Bob now have a consistent view of the chain.}
\label{fig:sc_reorg2}
\end{figure}

This scheme allows for powerful organic reorganizations to occur, as shown in Figures~\ref{fig:sc_reorg1} and~\ref{fig:sc_reorg2}.
For illustrative purposes, suppose there are two miners, Alice (blocks superscripted with $A$) and Bob (blocks superscripted with $B$).
In Figure~\ref{fig:sc_reorg1}, Alice and Bob each mine on top of the longest parent chain they are aware of; in this case, there is a tie, so they are each mining on a different fork.
By luck, Alice finds her block first and broadcasts it to the network.
Bob, seeing this new block, then begins mining on top of this new longest chain as shown in Figure~\ref{fig:sc_reorg2}.
This scheme supports organic short reorganizations, as each fork is internally consistent and locally canonical to the miner working on it.
This also allows the side chain to seamlessly support persistent chain split scenarios, \textit{e.g.}, in the case of a contentious hard fork.

\subsection{Security Analysis}
\label{sec:method:security_analysis}

The two-way peg provided by the proposed side chain scheme is trust-minimized.
We assume adversaries that can are weaker than a majority block producer censorship attack on the parent chain.
As our proposal relies on fraud proofs, this is an optimal adversarial model.

As merged consensus is permissionless and performs leader selection entirely on-chain, a side chain with merged consensus is no less available than its parent chain.
The proposed side chain scheme is deterministic, objective, and uses no off-chain information, so it is no less consistent than its parent chain.

Under the given adversarial model, anyone may become a side chain block producer, even if all other side chain block producers are censoring their transactions, so all state elements can be consumed subject to a finite delay \textit{i.e.}, they are live.
Anyone may fully validate the side chain client-side, as all data is available, and submit a fraud proof.
Unless the parent chain is being censored for the duration of the side chain block finalization delay (which is outside our adversarial model), such fraud proofs can be submitted, therefore state elements are safe.

The proposed scheme is therefore optimally trustless modulo an assumption on the small amount of block space needed to post a fraud proof.
This is an even stronger guarantee than what is provided by \textit{e.g.}, Plasma chain (\S~\ref{sec:related:plasmachains}) or channels (\S~\ref{sec:related:channels}), which are vulnerable to chain congestion as well and require lots of cheap block space.

\subsection{Further Improvements}
\label{sec:method:extensions}

The proposed side chain scheme can be further improved for both client-side performance and parent chain optimization.

\subsubsection{Parallelizable Data Authentication and Availability}

The primary scaling avenue for the proposed trust-minimized side chain design is the separation of consensus on execution from data for parent chain full nodes.
The authentication \textit{i.e.}, Merkleization, and broadcasting (transmission) of data is a stateless process however.
Transactions can be modified to flag an invariant: that certain parts of the transaction data are not be be processed by---or even accessible to---the chain's virtual machine environment, but are instead pre-processed by specified pure functions (such as, but not limited to, Merkleization), with only the results accessible to the execution environment~\cite{multi_threaded_data_availability}.

\subsubsection{UTXO-specific Fraud Proofs Without Intermediate State Serialization}

The general-purpose fraud proofs scheme in~\cite{fraud_proofs} can be used for \textit{any} computational model (and so can the side chain scheme presented in this work), but is not the most effective for client-side validation in all cases.
It involves serializing the state after every few (or every) transaction in order to compute a new intermediate state root; this is a very expensive process that is also a single-threaded bottleneck.

For the UTXO data model specifically, which is sufficient for a decentralized payment system, block producers can attach metadata to each input that commits to a claim on the exact output that it is spending.
If this claim is invalid, it can be proven with a non-interactive fraud proof.
This scheme does not require any intermediate state serialization, and in fact no state serialization at all~\cite{bip141,compact_utxo_fraud_proofs}.

\subsubsection{The State-lookupless Client Paradigm}

State accesses are the primary bottleneck for blockchains with stateful smart contracts, \textit{e.g.}, Ethereum.
The stateless client paradigm~\cite{utreexo} attempts to remove this bottleneck by instead having transactions include a \textit{witness} to the pre-state of the transaction along with the post-state elements, with full nodes only needing to store a logarithmic- or fixed-sized accumulator.
The issue with this approach is twofold: 1) light clients now need to rely on service providers to a greater extent, as they cannot craft a complete transaction without knowing the state and 2) witnesses are immediately outdated and so must be kept up to date, potentially increasing computation or network bandwidth requirements.

The state-lookupless client paradigm~\cite{state_lookupless} also allows transactions to provide a witness against a dynamic accumulator~\cite{smt} of the state.
Full nodes then also check this witness against some of the most recent blocks (a system parameter); if the witness is too old the transaction is invalid.
Since the state transitions of each transaction is uniquely and totally defined in the UTXO data model, the process of validating a stateless transaction, then ensuring it is not spending a spent output in subsequent blocks, is an entirely stateless process.
Finally, if valid and included in a block, the transaction's state transition is applied to the state, which must be kept by all full nodes.

\subsubsection{On-Chain Data Availability Proofs}

Rather than posting all data on-chain all the time, the data availability scheme of ~\cite{fraud_proofs} can be used.
However, as it relies on client-side randomness and consensus support, it cannot be implemented entirely on-chain, and must instead of exposed through a Foreign Function Interface~\cite{non_interactive_data_availability_proofs}, such as a precompile.
This is the core idea of blockchains such as LazyLedger~\cite{lazyledger}, which completely separate consensus on execution from data availability and ordering, allowing them to achieve the scalability of sharded systems without the complexities of sharded systems.

\subsubsection{Halting for Weaker Synchrony Requirements}

One potentially problematic feature of the proposed scheme, along with Plasma chains (\S~\ref{sec:related:plasmachains}), channels (\S~\ref{sec:related:channels}), and other scaling proposals that rely on fraud proofs, is that users of the side chain must be online periodically for as long as they have funds on the side chain (potentially forever).

Instead, we can simply \textit{halt} the side chain after a fixed, predetermined amount of time \textit{e.g.}, measured in blocks~\cite{side_chains_halting}.
Then we allow a very long period of time---potentially months---during which anyone may submit a fraud proof to roll back the tip of the chain, but not extend it with new blocks.
Only after this time has passed are side chain blocks considered finalized, and withdrawals can be performed non-interactively against the final state of the side chain.
Users now know exactly when they have to be online to validate the chain: a known, finite time slot.
If users wish to withdraw their funds early from the side chain, they may do with trustlessly via atomic swaps~\cite{atomic_swap} with a liquidity provider.

\section{Related Work}
\label{sec:related}

A wide range of scaling techniques have been proposed over the years, and are discussed in this section.
More importantly, an analysis of incentives and shortcomings for each of these techniques is shown.

\subsection{Validity Proofs and Succinct Arguments of Knowledge}
\label{sec:related:validityproofs}

Recent years have seen the emergence of almost-practical constructions employing succinct arguments of knowledge~\cite{zksnarks,zkstarks} that are zero-knowledge.
This class of protocols allows a prover to generate a proof of an arbitrary arithmetic circuit's correct execution over some input that can then be verified efficiently.
It may initially seem that using these is superior to constructions that make use of fraud and data availability proofs in the context of layer-2 scaling techniques, as the latter rely on an assumption that the parent chain is readily available to post challenges to while the former always guarantees correct state execution.
Unfortunately, circuit-based zero-knowledge protocols have fundamental limitations that make them inappropriate for use as a core component to scaling techniques.

First, proof generation is monopolistic rather than competitive as with PoW mining.
Mining is a random process~\cite{bitcoin}, and even a miner using pen-and-paper is capable of producing a block today if they get lucky; censoring other block producers requires a \textit{majority} of mining power.
Proof generation for these zero-knowledge protocols on the other hand is monopolistic: the user with the lowest-latency prover will always win the race to generate proofs first when attempting to prove execution of the same circuit over the same inputs.
Such a system tends towards becoming permissioned over time, especially when incentives for dispersing proving power are non-existent, and will resemble single-operator Plasma chain constructions in this regard---though without the exit game and synchrony assumptions needed by those (\S~\ref{sec:related:plasmachains}).

Second, and more importantly, a completely transparent blockchain or layer-2 system can be rolled back in the event of an implementation bug---either with a forced re-organization~\cite{bitcoin_rollback,bitcoin_rollback_cve} or a forced special state transition to revert unwanted effects~\cite{thedao_hack_fix}.
In contrast, in a system employing a circuit-based zero-knowledge protocol without full data availability with no further checks, a bug in either the implementation of the circuit or the trusted setup (if the protocol requires one)~\cite{zcash_bug} may result in \textit{permanent} state corruption~\cite{zcash_bug_turnstile} that cannot be recovered from save for restarting the chain from genesis.
For layer-2 constructions that have full data availability and use the zero-knowledge protocol only for proving correct execution of state transitions~\cite{roll_up}, a larger surface for implementation bugs exist, as off-chain code must be implemented correctly in addition to the on-chain smart contract that verifies proofs.
This is an especially egregious problem given the complexity of implementing arithmetic circuits and the current lack of mature tooling (\textit{i.e.}, formal verification, linting, etc.) for developing such programs.
Attempts to alleviate this make the use of zero-knowledge proofs redundant, and reduce to a Merkle computer verification game~\cite{truebit,arbitrum}, a Plasma chain construction~\cite{plasma}, or something similar.

\subsection{Merged Mining}
\label{sec:related:mergedmining}

\begin{figure}[!h]
\centering
\begin{tikzpicture}[every node/.style = {shape=rectangle, draw=black, align=center, minimum size=1cm}]
\node (ib) [draw opacity=0] {\dots};
\node (b1) [right=of ib] {$B_{i}$};
\node (b2) [draw opacity=0, right=of b1] {};
\node (b3) [right=of b2] {$B_{i+1}$};
\node (b4) [right=of b3] {$B_{i+2}$};
\node (b5) [right=of b4] {$B_{i+3}$};
\node (b6) [right=of b5] {$B_{i+4}$};

\node (is) [draw opacity=0, above=0.5cm of ib] {\dots};
\node (s1) [right=of is] {$S_{j}$};
\node (s2) [right=of s1] {$S_{j+1}$};
\node (s3) [right=of s2] {$S_{j+2}$};
\node (s4) [draw opacity=0, right=of s3] {};
\node (s5) [right=of s4] {$S_{j+3}$};
\node (s6) [right=of s5] {$S_{j+4}$};
\path[->]
    (b1) edge (ib)
    (b3) edge (b1)
    (b4) edge (b3)
    (b5) edge (b4)
    (b6) edge (b5)
    ;
\path[->]
    (s1) edge (is)
    (s2) edge (s1)
    (s3) edge (s2)
    (s5) edge (s3)
    (s6) edge (s5)
    ;
\path[->]
    (b1) edge (s1)
    (b3) edge (s3)
    (b5) edge (s5)
    (b6) edge (s6)
    ;
\end{tikzpicture}
\caption{Merged mining.}
\label{fig:merged_mining}
\end{figure}
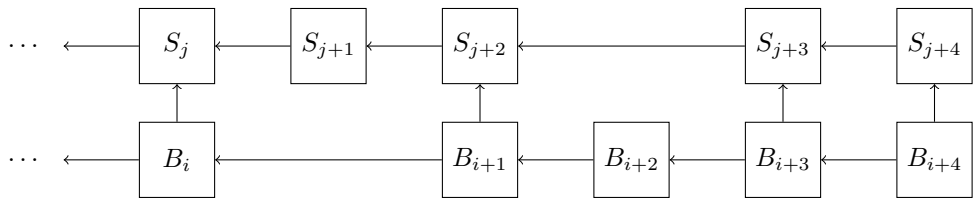

Merged mining~\cite{merged_mining_namecoin,merged_mining_forum} is a means of re-using computational power across two or more chains.
In order to merge mine a side chain with a parent chain, the block hash of a side chain block is included in a standardized way in the currently mined parent chain.
If the block satisfies the difficulty of either chain (with the side chain traditionally having lower difficulty) then it is considered a valid proof of work for that chain, and the block is appended to the appropriate chain~\cite{alternative_chain}.
This is illustrated in Figure~\ref{fig:merged_mining}, with some blocks of the parent chain ($B$) including hashes of the merge mined side chain ($S$).

Note that, when implemented with a na\"{i}ve longest-chain fork choice rule, this allows the side chain to re-use hashing power from the parent chain, but not borrow security.
Since parent chain blocks cannot be used for checkpointing in this scheme (especially if the difficulty of the side chain is higher than that of the parent chain), as parent chain blocks are merely superblocks~\cite{nipopow}, the merge mined side chain only gains security against \textit{external} hashing hardware~\cite{sia_asics}.

As a consequence of this, a common criticism against merged mining is the virtually zero cost of attacking a side chain by the miners of the parent chain, which does not apply to the scaling solution presented in this work.

\subsection{Pegged Side Chains}
\label{sec:related:sidechains}

The general idea of using side chains to deploy innovations and improvements to a chain without interruption has been suggested for many years~\cite{sidechains}.
Namecoin~\cite{namecoin} is one of the more prominent and early examples of a side chain that runs alongside Bitcoin and acts as a decentralized DNS.

\begin{definition}
A \textit{side chain} is a blockchain that validates data from one or more other blockchains (adapted from~\cite{sidechains}).
\end{definition}

In plain English, a side chain runs alongside a parent chain (or possibly more than one parent chain, though this configuration is not used much in practice) and ``understands'' the existence of a canonical parent chain.
This allows it to \textit{yank} data (events) from the parent chain to perform actions on its own state.
Note that there are no requirements on how a side chain is secured---indeed, running a side chain with its own independent consensus protocol is generally counter-productive as this will make it less secure than its parent chain.

It has generally been understood that a completely trustless and secure two-way peg of assets is impossible~\cite{drivechain}, though moving assets from the parent chain to the side chain is possible using the yanking scheme described above.

While there have been attempts to implement a two-way peg using light-client proofs~\cite{pos_sidechains,pow_sidechains,drivechain}, such constructions are vulnerable to a minority of block producers on the parent chain or a majority of block producers on the side chain---which presumably will be less costly to attack than the parent chain.

\subsection{Plasma Chains}
\label{sec:related:plasmachains}

Plasma~\cite{plasma} introduced Plasma chains as a potential scaling methodology.
At a high level, a Plasma chain operates in much the same manner as a side chain: funds (or, more generally, state), can be yanked from the parent chain---Ethereum---to the Plasma chain, while state can be \textit{exited} through a commit-challenge scheme known as an exit game.

An \textit{operator} is usually responsible for collecting transactions into blocks and committing block hashes to the parent chain (this allows the Plasma chain to borrow security from the parent chain without having to run a permissionless consensus protocol of its own).
Several variations of Plasma chain constructions have been proposed, using different data models~\cite{plasma_mvp,plasma_cash}, though they are all with significant unresolved issues.
Fungible iterations of Plasma~\cite{plasma_mvp} require \textit{mass exits}, as there are no guarantees of Plasma chain liveness or safety, while non-fungible iterations of Plasma~\cite{plasma_cash} require maintaining an every-growing history of proofs, or posting linear-sized checkpoints on-chain.

Operators can generally misbehave in two ways: 1) censoring a user's transactions or 2) attempting to fraudulently exit state (\textit{i.e.}, assets) back to the parent chain.
Each of these are resolved by allowing users to 1) force a state transition on the Plasma chain by executing it on the parent chain or 2) prove an invalid state transition occurred on the side chain, on the parent chain (which can be done implicitly in the case of a mass exit as a response to block withholding by a malicious Plasma operator).

Plasma chains are dependent on an honest majority of block producers for state safety; additionally, one critical caveat is that a Plasma chain is only trustless under the strictly stronger assumption that block space is cheaply available on the parent chain to either force a valid state transition or challenge an invalid state transition in finite time.

\subsection{Channels}
\label{sec:related:channels}

Channels were first envisioned as payment channels~\cite{payment_channels} between two or more parties to allow them to exchange money almost instantly without waiting for transactions to be included into blocks on a blockchain.
More general-purpose state channels~\cite{state_channels,counterfactual} were later described as a mechanism for participants of the channel to agree on potentially arbitrary state rather than just payments.

A channel proceeds by unanimous agreement among a fixed set of channel participants to update its state.
This allows them to have instant finality, as any participant can close the channel by publishing the agreed-upon latest state to the blockchain.
A user that attempts to close a channel with an old state can be met with a challenge with a more recent state, which by definition is signed by all parties.

While the instant finality offered by channels is undoubtedly a significant advantage over side chains and Plasma chains, unanimous agreement has several drawbacks.
First, all channel participants must be online in order to sign and agree to a state update, and the set of participants is fixed at channel creation.
Second, there is no way to distinguish a user who lost their copy of the most recent state with a malicious user attempting to close the channel with an old state to their advantage.
As only the latest state is valid in channel schemes, users can only make \textit{copies} of their local state, not \textit{backups}---the two protect against fundamentally different classes of data failures, with copies being strictly less useful.

Payment channel networks~\cite{lighting_network} aim to alleviate the problem of having a fixed participant set by allowing agreement to take place atomically between users with bidirectional payment channels open between themselves.
The issues this introduces are legion, and enumerating them is outside the scope of this work.

Note that similarly to Plasma chains, channels can only be made to be trustless if block space is available on the blockchain they operate on, and an assumption on an honest majority of block producers for state safety.

\section{Conclusion}
\label{sec:conclusion}

In this work we introduce a blockchain scaling solution that is both secure and decentralized in practice, and allows for greater transaction throughput than conventional blockchain systems deployed today.
In addition, several terms that have emerged in common blockchain parlance are given proper definitions so as to enable and encourage collaboration without confusion.

\printbibliography

\end{document}